\documentclass[%
 aip,
 jmp,%
 amsmath,amssymb,
 reprint,%
]{revtex4-2}

\usepackage{graphicx}
\usepackage{dcolumn}
\usepackage{bm}

\preprint{ }
 \usepackage[version=4]{mhchem}

\usepackage[normalem]{ulem}
\usepackage{graphicx}
\usepackage{dcolumn,xcolor}
\usepackage{bm}

\begin{document}

\preprint{}

\title[Oxidation by direct laser write]{Burst-mode fs-laser direct writing for full-thickness oxidation of Ta thin films}
\author{
Lina Grineviciute$^{1,*}$, Hsin-Hui Huang$^{2,3,4*}$, Haoran Mu$^{2,4}$, Nguyen Hoai An Le$^2$, Andrew Siao Ming Ang$^{2}$, Dan Kapsaskis$^{2}$, Tomas Katkus$^2$, Saulius Juodkazis$^{2,4,5}$
}%

\affiliation{Center for Physical Sciences and Technology (FTMC), Savanoriu ave. 231, LT-02300,Vilnius, Lithuania}
\affiliation{Optical Sciences Centre, 
Swinburne University of Technology, Hawthorn, Victoria 3122, Australia}
\affiliation{Australian Research Council (ARC) Industrial Transformation Training Centre in Surface Engineering for Advanced Materials (SEAM), Swinburne University of Technology, Hawthorn, VIC, 3122, Australia}
\affiliation{Melbourne Center for Nanofabrication (MCN), 151 Wellington Road, Clayton, Vic 3168, Australia}
\affiliation{Laser Research Center, Physics Faculty, Vilnius University, Saul\.{e}tekio Ave. 10, 10223 Vilnius, Lithuania}


\thanks{*L.G. and H-H.H. contributed equally. Correspondence: lina.grineviciute@ftmc.lt; hsinhuihuang@swin.edu.au }

\date{\today}

\begin{abstract}
Direct fs-laser (1030~nm/200~fs) write of a throughout oxide \ce{Ta2O5} on a 200~nm Ta film was achieved using a combined ps- and ns- burst mode (Burst-in-Burst or BiB) of fs-pulse exposure at a high 0.6~MHz repetition rate. Few micrometers-wide lines were formed at the center of 12~$\mu$m focal spot by controlled oxidation without ablation. The oxidized regions were flat and optically transparent. Wavelength-scale self-organized ripples of oxidized \ce{Ta2O5} sub-1~$\mu$m gratings were recorded by rastering a $1\times 1$~mm$^2$ area. The oxidized ripples with periodic pattern $\sim wavelength$ were aligned with the polarization of the writing beam. Energy deposition in the burst-mode oxidation is discussed by comparing 200~fs and 20~ps BiB-mode writing modes. The presented strategy of self-guided oxidation with heat deposition by BiB fs-laser opens an opportunity for debris-free and annealing-free oxidation on a sub-wavelength scale.
\end{abstract}

\maketitle
\tableofcontents
\vspace{0.2cm}

\begin{quotation}
Direct write of transparent oxide through the sub-wavelength-thick metal film is achieved without any post-processing and ablation by bursts of fs-laser pulses. 
\end{quotation}

\section{\label{intro}Introduction}

Direct laser writing of functional patterns endowed with tailored optical, photo-catalytic, mechanical, thermal, and tribological properties on surfaces of different materials is highly promising for a wide range of applications. Laser ablation and surface self-organization are widely explored with pulsed lasers from nanosecond to increasingly more popular femtosecond (fs)-pulse durations for wettability/icing control~\cite{23aem2300575}, coloration, anti-reflection, and radiative cooling at IR wavelengths (IR black-surfaces)~\cite{25as2076}. Furthermore, useful material response can be harnessed at low pulse fluences below the ablation threshold, e.g., phonon generation (THz emission) from Bi at $\sim 7$~mJ/cm$^2$ by fs-pulses (775~nm/50~fs)~\cite{Eugene} or oxidation and coloration of metals by ns-pulses at fluences below evaporation threshold~\cite{Veiko,Jwad}. In the latter case, the color of stainless steel AISI 304 and Ti was dependent on the surface temperature ($1300-2800$K) and a cumulative exposure time (tens-of-$\mu$s) achieved by pulses of 10-30~GW/cm$^2$ intensity (1.06~$\mu$m/100~ns)~\cite{Veiko,Lu_2017,Zhou_2018a,Ivanova_2025}. While Long ns-laser pulses form micrometers-thick oxide films during polishing, such treatments have been found to reduce the mechanical strength of Ti aviation alloy is reduced~\cite{TA15}. Additionally, the oxidation of metallic multi-component high entropy alloys (HEAs) was found to decompose on the surface into oxide and spinel forms according to their cohesive energies and favorable thermodynamic conditions according to the Gibbs energy~\cite{Wei}. Beyond bulk alloys, substrate temperature during pulsed laser oxidation remains a decisive factor, as evidenced in the growth of \ce{Ta2O5} films~\cite{Atanassova_2006}. Developing an atomistic-scale understanding of these subsurface mechanisms is essential, as they dictate the performance limits of Ta-based thin-film electronic devices~\cite{Kingon_2000}. Recently, coloration of Ti was made with fs-laser in burst mode and showed reduced formation of ripples~\cite{Geng}.

Oxidation of Si by slow scan of fs-laser beam with subsequent KOH removal of Si beneath the pattern was used to make suspended micro-spiral~\cite{Lei}. 
Dynamic surface oxidation of Si for dry etching mask application has also been demonstrated via seeding-oxidation, when an initial multi-pulse exposure is used to induce structural defects to increase absorption (at 126~mJ/cm$^2$ fluence), after which the oxidation front is dynamically controlled by fs-laser beam scan at a lower fluence 48~mJ/cm$^2$~\cite{YIN}. The controlled formation of small nanoscale features, e.g., nano-bumps and single nano-grooves were always at the focus of well controlled formation required in a number of applications, and fs-laser irradiation is an essential tool to deliver such nano-structures by surface melting with fast quenching on glass~\cite{03apl2901} or ablation near threshold~\cite{13ome1674}. Another example of fs-laser oxidation (not ablation), which was kept within the focal spot of $2r\simeq 1~\mu$m width using 400~fs/515~nm scanned laser pulses, is burning of a crystalline diamond via a graphitization step into \ce{CO2} for inscription of Fresnel micro-lenses without ablation debris~\cite{23ol1379}. Only 10~nJ pulses at 3.18~TW/cm$^2$ average intensity (or 1.27~J/cm$^2$ pulse fluence) were required under multi-pass scan conditions at 0.1~MHz laser repetition rate; the ablation threshold of diamond for the same pulse duration is 10~J/cm$^2$~\cite{diamo}. This removal of diamond by oxidation was confined within the depth close to the Rayleigh length of $\pi r^2/\lambda \sim 1.2~\mu$m at a single focus during laser scan. 
\begin{figure*}[tb]
\centering\includegraphics[width=.99\textwidth]{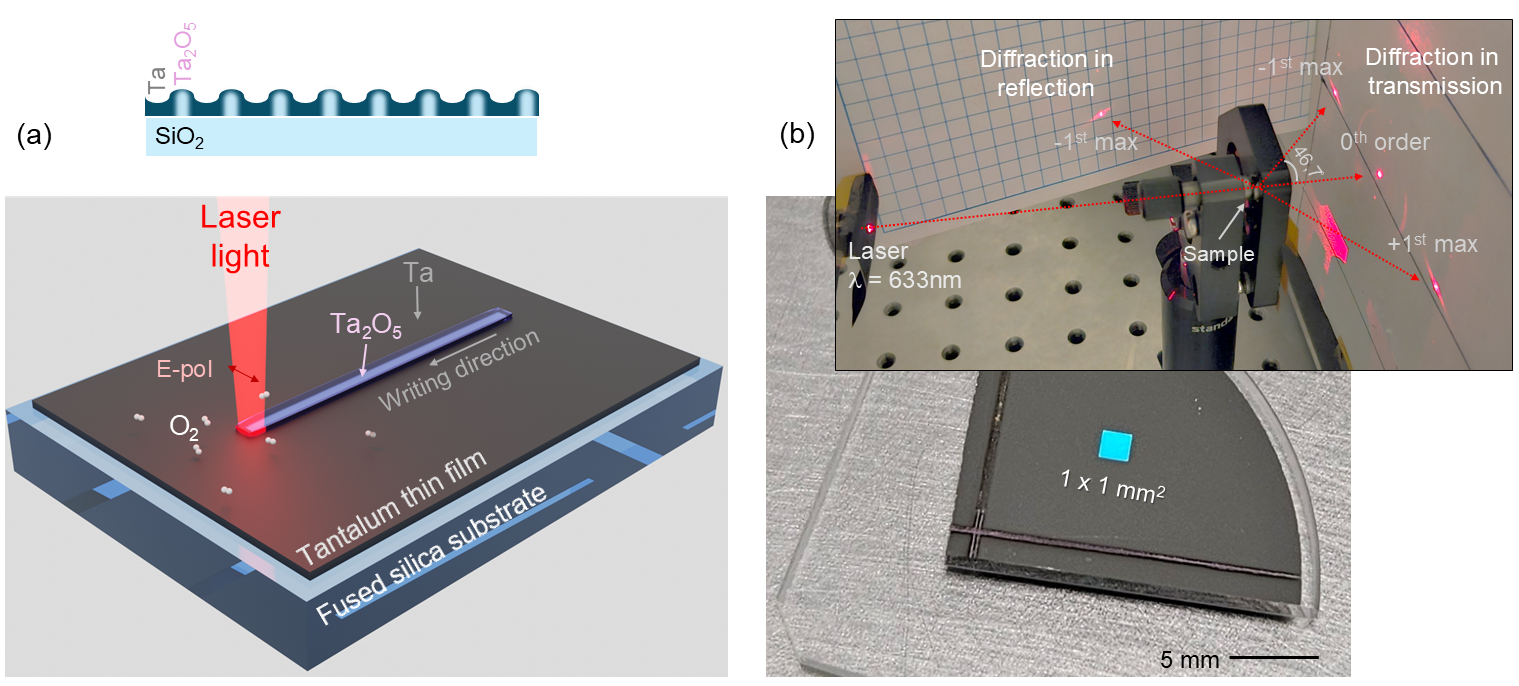}
\caption{\label{f-nice} Concept: (a) fs-laser driven oxidation of metal in the Burst-in-Burst mode via direct write at (b) sub-diffraction limit on micrometer scale and self-organization of ripples at sub-wavelength scale. 
The Transparent-Opaque patterning is achieved under ambient conditions. 
Diffraction on a laser-inscribed ripples' grating of $\Lambda = 860$~nm period, where transparency was achieved as a result of complete tantalum oxidation.}
\end{figure*}

Femtosecond laser field is currently driven by the demand for higher average power to meet industrial requirements for cutting, drilling, and dicing. This tendency follows the optical Moore's law over the last 25 years, where the combination of a higher laser repetition rate $f_l\sim 1$~MHz and a larger pulse energy $E_f\sim 1$~mJ delivers average powers in $\sim 1$~kW range~\cite{21ep44}. High average power is achieved in modern ultra-short pulse lasers by utilizing a burst mode operation, which allows to extract more energy from the laser crystal. While laser machining with fs-lasers in burst mode delivers shorter processing time for material removal, it also introduces a trade-off in the quality and making the underlying mechanisms a subject of intense research~\cite{Zema,ZEMAI,Yang_2024}. 
Burst mode employed needs fine-tuning for specific tasks base on the materials and other processing parameters such as ablation threshold, residual heat accumulation, or the specific temporal length of the burst envelope~\cite{Omer,Bonamis_2019,Forster_2021,Yang_2024}. While optimizing these parameters for material removal, Burst-in-burst (BiB) mode was found a particularly versatile method for surface modification, such as the precision coloration of stainless steel as compared with MHz and GHz bursts~\cite{Gaidys_2023,GAID}. A pioneering burst mode study hinted that the ablation efficiency increases significantly at lower pulse energies, as rapid pulse intervals outpace thermal diffusion to localize energy more effectively~\cite{Omer}. 

The current study utilizes this temporal energy-stacking combining high repetition rates with BiB-mode with low pulse energies comparable to those used in laser 3D polymerization. Under this condition, the surface temperature is sustained in a regime that facilitates oxygen diffusion into the lattice while remaining below the ablation threshold. Here, a direct write oxidation of a Ta 200~nm film on \ce{SiO2} substrate is demonstrated using a BiB-mode of fs-laser exposure (Fig.~\ref{f-nice}). Such modification renders a metallic light-blocking surface into a highly transparent $\sim 4$~eV semiconductor with promising optical nonlinearity and photo-chemical electrode applications~\cite{Ren}. 
Notably, this direct write oxidation of metal thin film is achieved at sub-diffraction resolution and feature size. 

\section{Samples and Methods}

Thin 200~nm ion beam sputtered (IBS, Cutting Edge Coatings GmbH) Ta films on fused silica glass were used in this study. Tantalum is a transitional and refractory element with one of the highest mass densities $\rho_0 = 16.7$~g/cm$^3$ and a high melting point of 3017$^\circ$C~\cite{PubChem_Ta_2025}. Melting temperature of \ce{Ta2O5} is considerably at 1872$^\circ$C~\cite{Waring_1968}. Oxidation of Ta to the final \ce{Ta2O5} state proceeds via sub-oxides~\cite{oxid}; an annealing of 1~hour at 600$^\circ$C was sufficient to transform the entire 200~nm film to \ce{Ta2O5} \cite{Nikitina_2024}. Refractive index of Ta at the wavelength of laser processing $\lambda = 1030$~nm is $n+i\kappa = 0.97326 +i9.5129$~\cite{Cheikh}, the skin depth $l_{s} = \frac{c}{\omega\kappa} = \frac{\lambda}{2\pi\kappa}\simeq 17$~nm (E-field penetration) and twofold lower for the intensity $I\propto E^2$.
\begin{figure*}[tb]
\centering\includegraphics[width=1\textwidth]{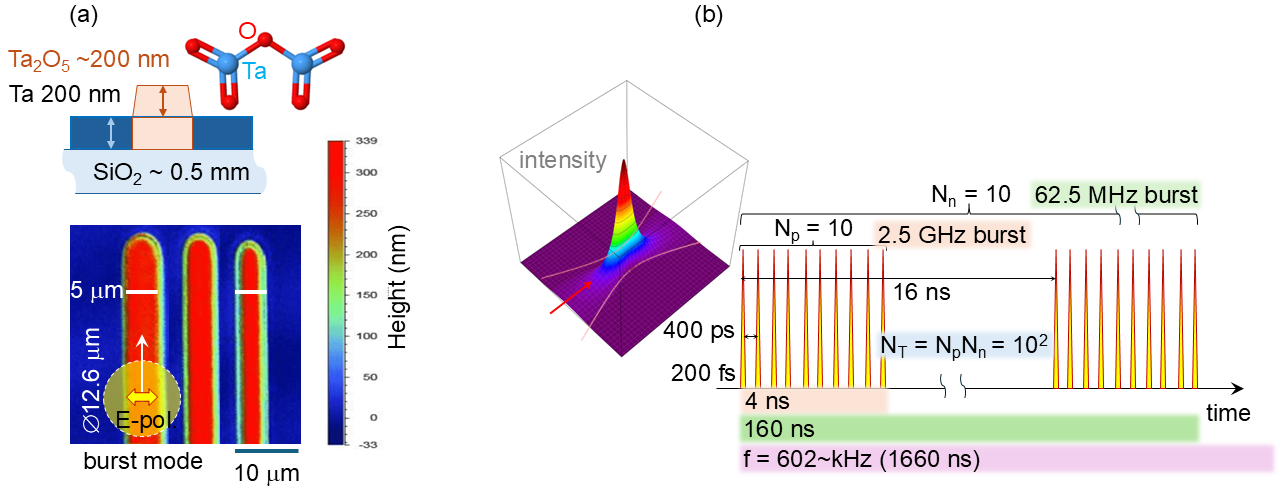}
\caption{\label{f-intro} Burst oxidation of Ta (Burst-in-Burst or BiB-mode). (a) Schematics of sample's cross section and an actual profile (Bruker, Germany) of the oxidized \ce{Ta2O5} sub-diffraction-limited micro-lines formed by scan of a $1.22\lambda/NA =12.6~\mu$m diameter focal spot (the wavelength $\lambda = 1030$~nm, numerical aperture of objective lens $NA = 0.1$); single fs-pulse energy for BiB was $E_f\simeq 0.97$~nJ (on the sample) and the conditions of exposure are discussed in the text. Oxide thickness is twice the original Ta 200~nm film (see Fig.~\ref{f-side} for cross section). (b) The used BiB mode (Carbide, Light Conversion) of 1030~nm/200~fs pulses: number of pulses in ps-burst-1 $N_p = 10$ (2.5~GHz burst) and in ns-burst-2 $N_n = 10$ (62.5~MHz burst) at an overall 602.7~kHz repetition rate. }
\end{figure*}

The following methodology was used to determine the average fluence and intensity of a single 1030~nm/200~fs pulse at laser (40~W Carbide, Light Conversion Ltd.) repetition rate $f_l = 602.7$~kHz from average power $P_{av}$. Single fs-pulselet energy $E_{f} = \frac{P_{av}}{f_l}\times\frac{1}{N_pN_n}\equiv 10^{-2}\times\frac{P_{av}}{f_l}$, since we used maximum number of pulses in ps and ns bursts $N_p=N_n=10$ (Fig.~\ref{f-intro}(b)). The average intensity of fs-pulse was calculated as $I_{f} = \frac{F_{f}}{t_{f}}$, where $t_{f} = 200$~fs is the single fs-pulse duration, $F_{f} = E_{f}/(\pi r^2)$ is the corresponding fluence where the radius of the focal spot 
$r = 0.61\lambda/NA$ defined by the wavelength $\lambda = 1030$~nm and the numerical aperture $NA = 0.1$ of the objective lens.

For example, at $P_{av} = 100$~mW in BiB-mode, the single fs-pulse energy $E_f = 1.66$~nJ, $F_f = 1.34$~mJ/cm$^2$ and $I_f = 6.69$~GW/cm$^2$, the intensity which is comparable with metal coloration by ns-pulses~\cite{Veiko}. Such fs-pulse fluence is approximately $2\%$ of single pulse ablation, while few-nJ pulse energies are typical for 3D polymerization by fs-pulses. Polarization was chosen linear (perpendicular to the scan direction) or circular (indicated where it matters). Beam scanning speed was $v_s = 10 - 100~\mu$m/s (over 5 lines of pattern). Time of BiB-mode energy deposition by 100 fs-pulslets $N_p\times N_n$ is defined by arriving energy $E_{bb} = P_{av}/f_l$ and number of exposure events (BiB) over the dwell time while laser beam is passing the focal diameter $t_{dw} = d/v_s$; the focal diameter $d \equiv 2r = 1.22\lambda/NA$. The total cumulative exposure dose by BiB-mode is $D_{\Sigma} = E_{bb}\times\frac{1}{\pi r^2}\times(t_{dw}f_l)$~[J/cm$^2$], where $N=t_{dw}f_l$ is the number of exposure cycles based on laser repetition rate (each cycle has the same amount of bursts $N_pN_n$). These parameters can be compared with single fs-pulse exposure in terms of fluence, intensity, and dose. Also, longer pulse durations of 20~ps were tested for DLW oxidation (indicated where it applies). 

For sub-wavelength oxide pattering by ripples, a single fs-pulse exposure was used with raster scanning. The period ripples were $\Lambda\simeq\lambda$ and the width of a bulging out oxide structure was $\lambda/2$. 

\section{Results}

The hypothesis of direct laser writing (DLW) of metal oxides throughout the entire layer without ablation was tested in this study. Tantalum was chosen due to the high optical transmission and nonlinearity of \ce{Ta2O5}. The aim was to define oxide structure purely by DLW without the required steps of additional annealing or chemical processing, which would reduce the applicability of the process. It is known that Ta is oxidized in air starting from modest temperatures $>300^\circ$C~\cite{oxid} and proceeds via increasing oxidation state~\cite{Kofstad_1961,Demiryont_1985}.

\subsection{Sub-diffraction limited width of \ce{Ta2O5} lines on Ta}

Tantalum is a good electrical and thermal conductor. Hence, a first harmonic 1030~nm/200~fs radiation was chosen to facilitate free carrier absorption, which scales as $\propto\lambda^2$. To improve energy deposition, a high repetition rate of 0.6~MHz and burst-in-burst (BiB) mode was selected. The maximum number of ps-burst of $N_p = 10$ with $\Delta t_p = 400$~ps between fs-pulslets and ns-burst $N_n = 10$ with $\Delta t_n = 160$~ns was selected. Single fs-pulse ablation threshold is approximately $\sim 0.1$~J/cm$^2$ for most metals~\cite{Chichkov_1996,Nolte_1997,Mittelmann_2020}, a value consistent across various transition metals due to the localized nature of ultrashort energy deposition~\cite{Hashida_2001}. Only $\sim 1\%$ of such a threshold was used in this experiment, expecting a controlled heat accumulation and oxidation. 
\begin{figure*}[tb]
\centering\includegraphics[width=1\textwidth]{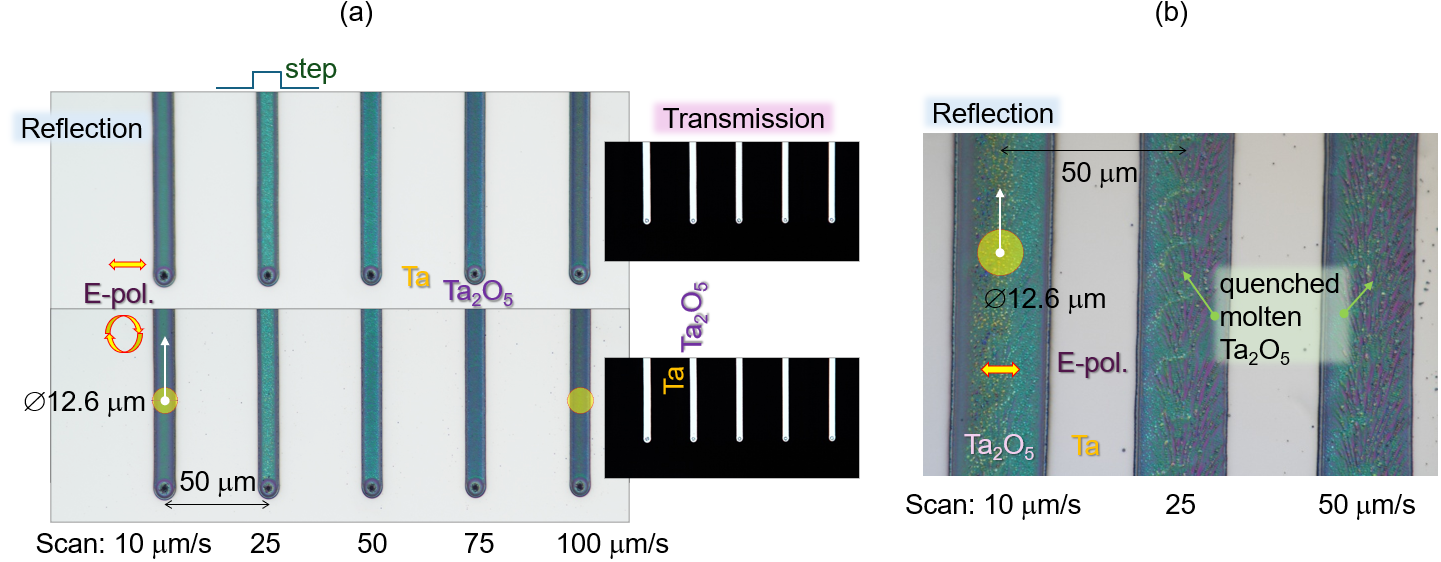}
\caption{\label{f-step} (a) Optical images in reflection and transmission of BiB-mode oxidized Ta film on \ce{SiO2}. Pulse energy of a single fs-pulslet in BiB was $E_f = 1.35$~nJ. Line exposure was carried out in the constant frequency mode, and the starting location received larger exposure, which caused ablation. (b) Wider than focal spot oxidation of Ta with apparent molten phase thermally quenched; full power corresponding to fs-pulslets $E_f = 14.51$~nJ.        }
\end{figure*}
\begin{figure*}[tb]
\centering\includegraphics[width=1\textwidth]{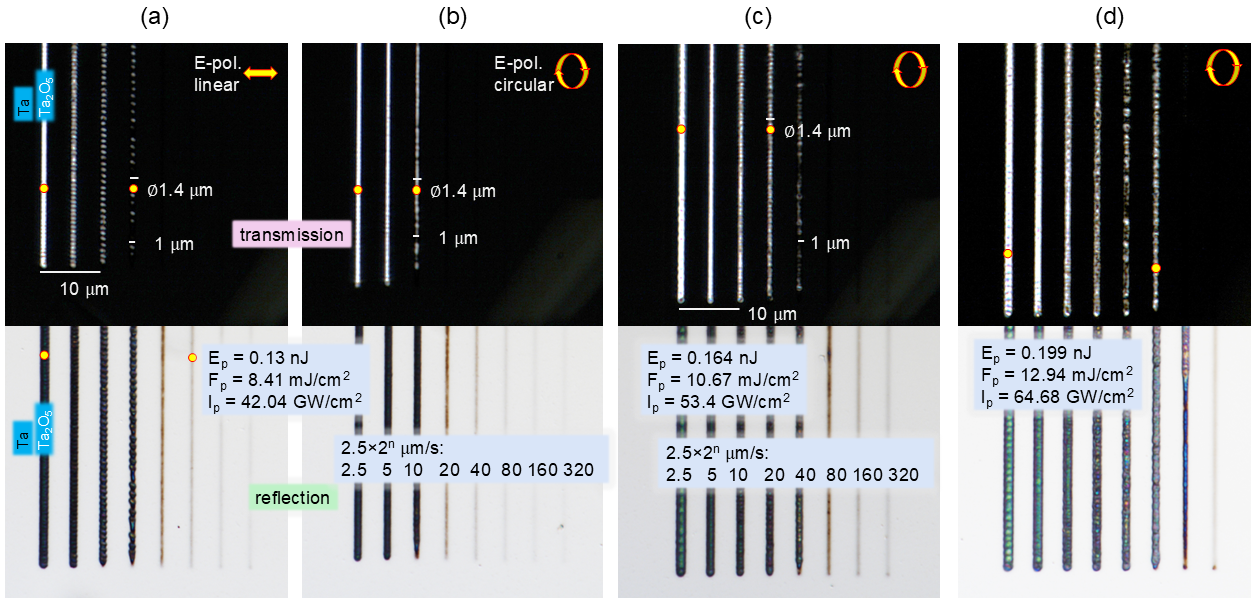}
\caption{\label{f-09na} Direct write oxidation using $NA = 0.9$ objective lens. Transmission and reflection optical images of the laser oxidised lines at different scan speeds and pulse energies for linear (a), and circular (b-d) polarizations. Focal diameter $1.22\lambda/NA = 1.4~\mu$m; scan speed was scaled as $2.5\times 2^n~\mu$m/s with $n = 0-7$. The number of BiB combined pulses per diameter of focal spot at the laser repetition rate $f_l$ is $N = t_{dw}f_l =337.5$ to 2.6 pulses for the lines (from left to right).    }
\end{figure*}

\begin{figure*}[h!]
\centering\includegraphics[width=1\textwidth]{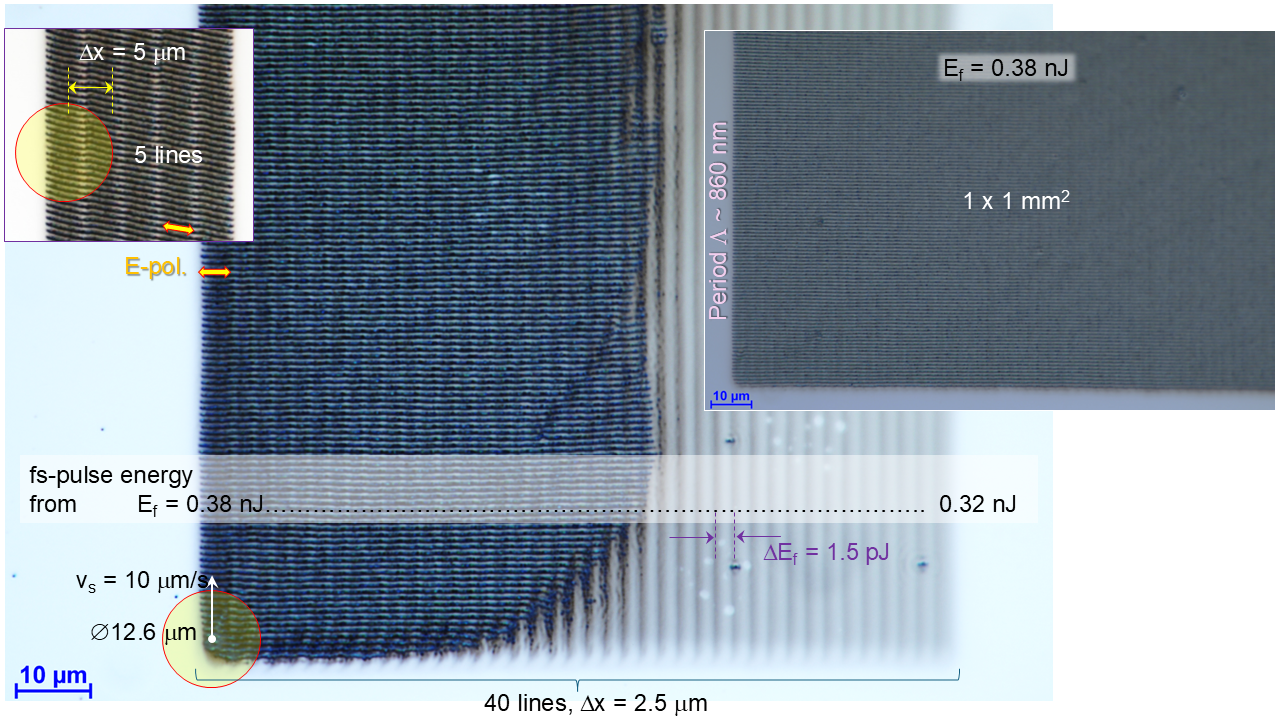}
\caption{\label{f-area} Optical images in reflection of surface patterns. Single-pulse oxidation of Ta at 602.7~kHz by self-organization of ripples. Focal spot of $1.22\lambda/NA =12.6~\mu$m diameter (the wavelength $\lambda = 1030$~nm, numerical aperture of objective lens $NA = 0.1$); single fs-pulse energy was changed from $E_f\simeq 0.38$~nJ to 0.32~nJ (on the sample) in 40 steps with energy reduction by $\Delta E_f = 1.5$~pJ during line-by-line scan with $\delta x = 2.5~\mu$m step between the lines. Scanning was one directional (arrow marker). Right-inset shows corner region of $1\times 1$~mm$^2$ area recorded in $\sim 10$~hours scan (see Fig.~\ref{f-nice}(b)). }
\end{figure*}
\begin{figure*}[tb]
\centering\includegraphics[width=1\textwidth]{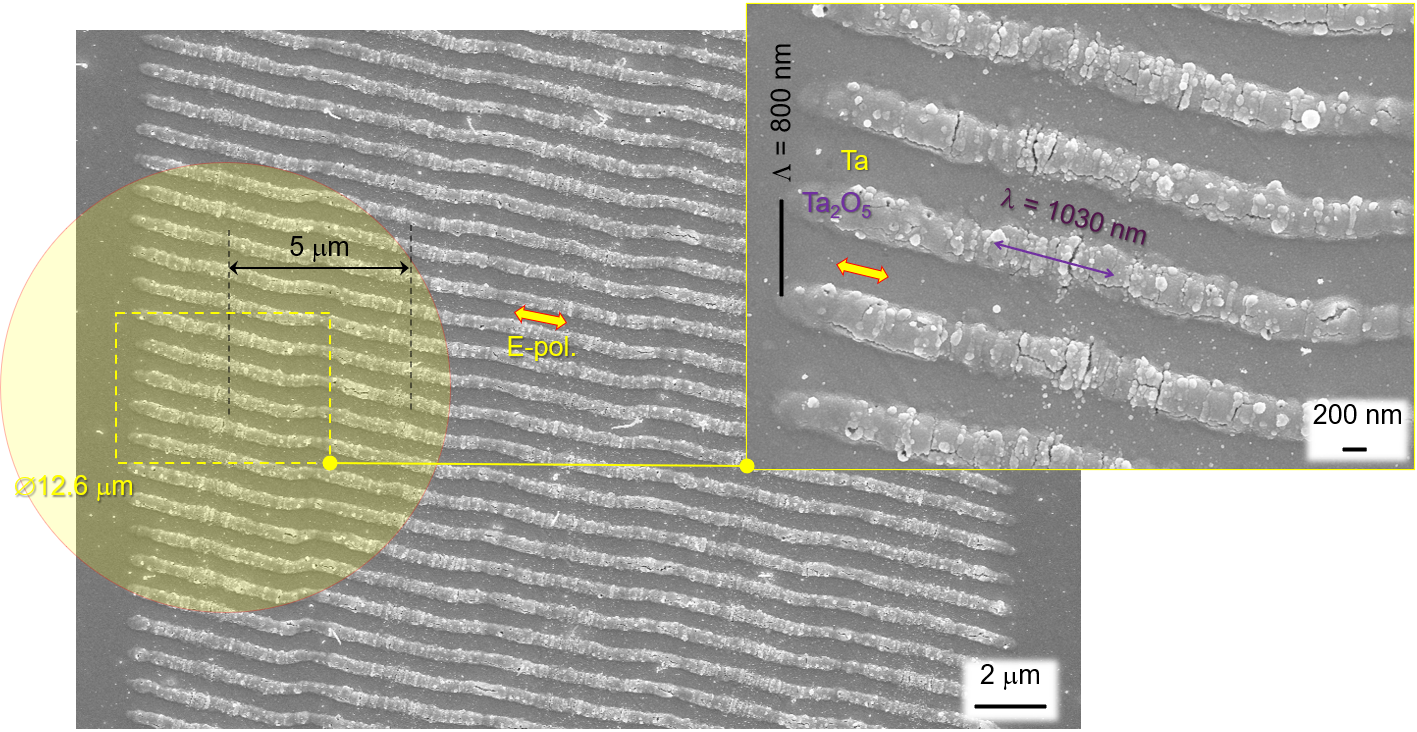}
\caption{\label{f-sem} SEM images of single-pulse oxidation of Ta at 602.7~kHz by self-organization of ripples (same area as in top inset of Fig.~\ref{f-area}). Focal spot of $1.22\lambda/NA =12.6~\mu$m diameter (the wavelength $\lambda = 1030$~nm, numerical aperture of objective lens $NA = 0.1$); single fs-pulse energy was $E_f\simeq 0.38$~nJ (on the sample) and line-by-line scan with $\Delta x = 5~\mu$m step between the lines. The right inset shows the zoomed-in region. }
\end{figure*}

Unique feature of the DLW driven oxidation of 200~nm Ta film was an overall flat top structure of $\sim 400$~nm tall \ce{Ta2O5} (Fig.~\ref{f-intro}(a)). Those three lines were formed in BiB-mode where the single fs-pulse energy was $E_f = 0.966$~nJ 
fluence per single fs-pulse $F_f = 0.78$~mJ/cm$^2$ and average intensity $I_f = 3.89$~GW/cm$^2$ at scan speed $v_s = 10, 25, 50~\mu$m/s (Fig.~\ref{f-intro}(a)). The total energy per all ps-ns bursts is $E_{bb} = E_f\times N_pN_n = 96.6$~nJ, and the total dose per focal diameter was $D_\Sigma = 59.2$~kJ/cm$^2$ for the slowest $v_s = 10~\mu$m/s. For the slowest $10~\mu$m/s scan the resulting \ce{Ta2O5} log was $\sim 5~\mu$m wide, while for the $50~\mu$m/s faster scan it was narrower by a factor of $\sqrt{5}\simeq 2.2$, which is due to diffusional scaling (heat diffusion length $\propto\sqrt{time}$); the height of all structures was the same. The entire oxidized region had a flat top and was approximately twice as tall as compared with the 200~nm initial Ta film (Fig.~\ref{f-intro}(a)). This is consistent with a height change observed in a thermally (homogeneously) oxidized Ta of the same thickness made in a separate experiment. 

The BiB driven oxidation of Ta has a self-limiting character. Once the entire 200~nm of Ta at the focal cross section is turned to \ce{Ta2O5}, it becomes transparent. Figure~\ref{f-step}(a) illustrates this condition. The lines oxidized at 10 times different scanning speed had the same width and height. Interestingly, there were no ripples formed in this BiB-mode of oxidation, and oxidized lines looked identical for linear and circularly polarised laser beam. More smooth (a single color in the reflection image) oxide lines were formed at a slower scan when a higher surface temperature was expected. At the highest fs-pulse energy in BiB-mode, the oxidized lines become wider than the focal spot and acquire a saturated height (Fig.~\ref{f-step}(b)). For the widening of \ce{Ta2O5} to occur, there should be a temperature spread from the original focal spot before the entire thickness of Ta is oxidized. Oxidation of the entire 200~nm of Ta occurred when the temperature exceeded 600$^\circ$C in air (annealing in oven). At high fs-pulselet energy $E_f \geq 12$~nJ, there was morphology consistent with molten phase movement as oxidation progressed during fs-BiB beam scan (Fig.~\ref{f-step}(b)). Melting of \ce{Ta2O5} requires high $\sim 1870^\circ$C temperature. The high surface temperature of Ta-to-\ce{Ta2O5} transformation during laser beam scan explains the absence of debris and ripples on the surface, which are useful virtues of this method of BiB-mode direct write of transparent oxide. At a faster scan $> 50~\mu$m/s, there was ablation and debris apparently due to larger thermal gradients and stress. Slow scan and BiB-mode were essential for controlled oxidation of Ta.    

Oxidised Ta lines were written at the highest dry objective lens $NA = 0.9$ (Fig.~\ref{f-09na}). The dwell time for laser to travel the distance equal diameter of focal spot $t_{dw} = 2r/v_s = 0.56$~s to 0.4375~ms, which at $f_l = 602.7$~kHz repetition rate defines number of pulses per focal diameter $N = t_{dw}f_l =337.5$ to 2.6 pulses. The cumulative dose was from $D_\Sigma = 283.8$~J/cm$^2$ to 2.29~J/cm$^2$ (a), 360.1-to-2.9~J/cm$^2$ (c), 436.7-to-3.52~J/cm$^2$ (d). When the cumulative dose $D_\Sigma \leq 35$~J/cm$^2$, there was no throughout oxidation of Ta (no transmission along the scanned line). The color appearance of scanned lines in reflection at lowest dose looked with color due to surface oxidation, which was much shallower than the 200~nm Ta film. Interestingly, circular polarisation was useful to obtain more homogeneous line along the scan (at the same power), however, a slightly larger cumulative dose was required at the threshold of formation for the oxidation line (a) vs. (b) in Fig.~\ref{f-09na}. This can be explained by peak electric field $E_p^{cir} = E_p^{lin}/\sqrt{2}$; the $\lambda/4$-plate is aligned with the fast- or slow-axis for the linear polarisation output and is at $45^\circ$ rotated for the circular. The intensity $I=|E|^2$ becomes twice larger for the linear as compared with circular polarisation. Similar dependence for multi-photon ionisation of \ce{SiO2} and \ce{Al2O3} was observed~\cite{Temnov}.

The ablation threshold intensity is reached in metal when the laser pulse's E-field driven electrons reach energy in excess of the electron work function and exceeds it by the binding energy of metal, i.e., departing electrons from the metal surface pulls ions out via the electrostatic mechanism~\cite{Eug}; see discussion of ablation in Sec.~\ref{abla}. This typically requires peak fluences above 100-200~mJ/cm$^2$ (note, we present results in average fluence in this study which is twice lower than peak for the Gaussian pulse profile).   

The lines which were not oxidised through the entire thickness of Ta nano-film, are clearly recognisable (Fig.~\ref{f-09na}) including the colar-dose dependence. Further studies are required to investigate nanoscale composition in those regions. It is also discernible that there was additional material distribution with micro-periodicity along the scanned line dependent on scanning speed. This phenomenon is linked to the combined visco-elastic surface flow and energy deposition which are affected by material property, local surface curvature and roughness, similar to the observation in fs-laser volume structuring, where periodic patterns are formed along the scanned line~\cite{Yves,05apa725} (detailed investigation of such structures was beyond the scope of this study).  

\subsection{Area pattering by sub-wavelength \ce{Ta2O5} ripples}

Self-organization of oxidized \ce{Ta2O3} ripples with surface expansion by $\sim 200$~nm from the original Ta surface were patterned over areas up to $1\times 1$~mm$^2$ by single 1030~nm/200~fs pulses at $602.7$~kHz repetition rate (Fig.~\ref{f-area}). For the $E_f = 0.38$~nJ single pulse, the fluence is $F_f = 0.31$~mJ/cm$^2$ and the average intensity $I_f = 1.53$~GW/cm$^2$. The cumulative dose over the focal spot (the dwell time) is $D_\Sigma = F_f\times\frac{d}{v_s}\times f_l = 232.37$~J/cm$^2$. Figure~\ref{f-area} shows a transition from a robust patterning at 0.38~nJ/pulse to no ripples at 0.34~nJ/pulse for the used 80\% overlap between the neighboring vertical scans. The period of oxidized ripples was $\Lambda = 800$~nm, and they are aligned with the polarization of the laser beam. This type of sub-wavelength ripples patterning was debris-free and is based on dipole nature of light scattering at the locations of initial energy deposition (absorption locations), which protrudes above the initial Ta surface (see discussion section~\ref{disco} and previous study where Ta was structured by 515~nm/200~fs pulses~\cite{26a}).
Large area patterning is possible by expanding the ripple pattern over required areas (see right-inset in Fig.~\ref{f-area}). To expedite such patterning, an obvious improvement is by a cylindrical lens. 

Figure~\ref{f-sem} shows an SEM image of \ce{Ta2O5} ripples in Ta 200~nm film. Surface swelling and oxidation induced cracking is apparent at the scale of tens-of-nanometers. Such nano-features are much smaller than the metal-dielectric grating of period $\Lambda \simeq 800-860$~nm.  

\begin{figure*}[tb]
\centering\includegraphics[width=1\textwidth]{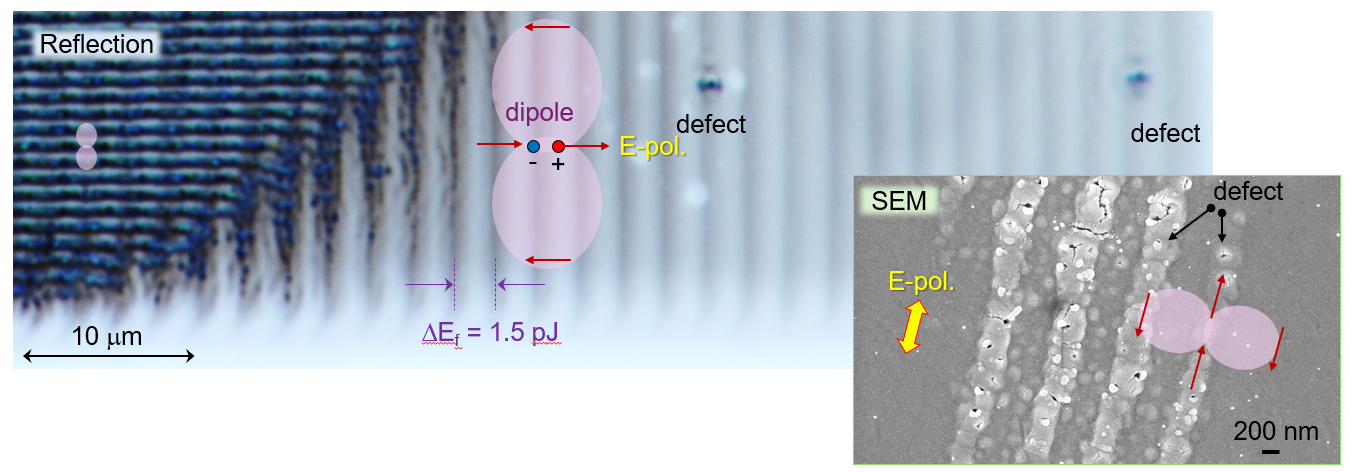}
\caption{\label{f-dipo} Dipole scattering at the defect site where energy is deposited, and oxidation starts, i.e., surface protrusion is formed. Optical image (same as Fig.~\ref{f-area}) shows darker regions with scattering signatures. SEM image is for a single vertical scan at a tilted (to scan direction) linear polarization; pulse energy $E_f\simeq 0.38$~nJ.        }
\end{figure*}

\section{Discussion}\label{disco}


\subsection{Ripples formation via dipole scattering}

Figure~\ref{f-dipo} shows close up view (optical image) of ripple formation at the beginning of the line, when pulse energy is gradually decreased by 1.5~nJ between the lines shifted by 2.5~$\mu$m (same sample as in Fig.~\ref{f-area}). The dipole-like pattern of light scattering is recognisable in reflection from the Ta surface, which is most apparent at the defect sites. Those locations of energy deposition or surface protrusions become scattering sites, which facilitate further energy deposition via interference from other regions of nanoscale protrusions. Those surface nano-features which are $\sim\lambda$ separated will contribute to a constructive interference, which leads to ripple pattern formation. Such a mechanism is established in ablation ripples and nano-gratings inside the bulk of transparent materials~\cite{Jean}.
SEM image (Fig.~\ref{f-dipo}) reveals the pattern of nano-seeds of \ce{Ta2O5}, which constructively couple when separation is $\sim\lambda$. 

\begin{figure*}[b!]
\centering\includegraphics[width=.9\textwidth]{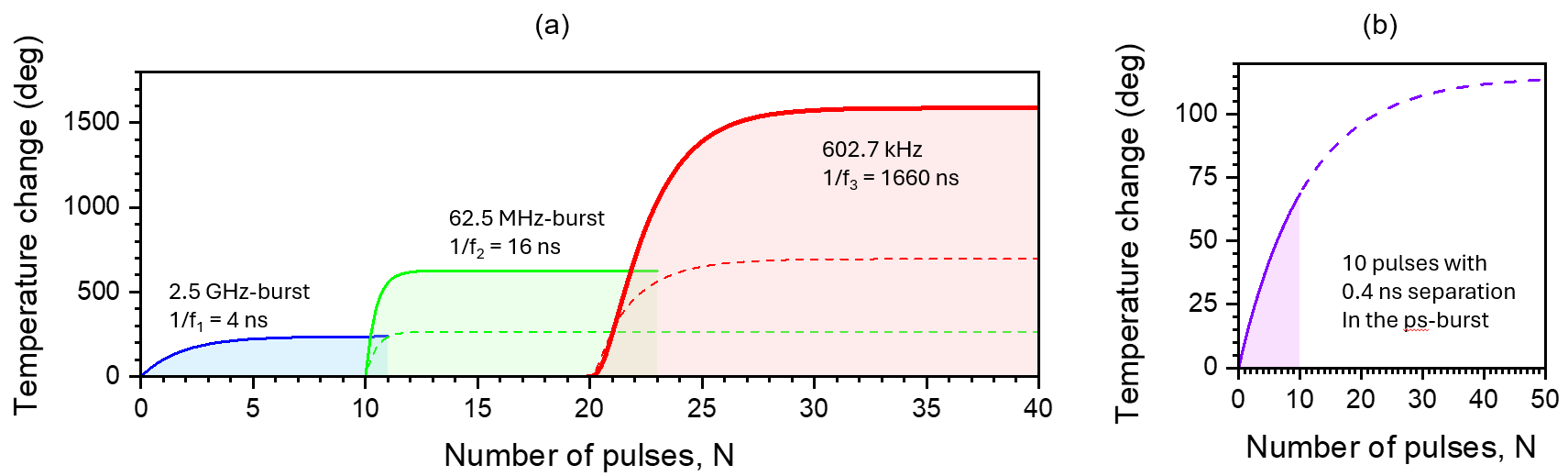}
\caption{\label{f-temp} Toy-model of heat accumulation. (a) First pulse heats focal spot by $\Delta T = 80^\circ$C (from RT 20$^\circ$C). The cooling time of the focal spot by heat diffusion at thermal diffusion of Ta $\alpha = 20$~mm$^2$/s is $t_{th} = h^2/\alpha = 2$~ns, where the thickness of Ta $h = 200$~nm (close to skin layer of light energy deposition $\lambda/4$). Temperature rise vs number of pulses $N$ is modeled~\cite{Gam} as $T_N = T_1\times\frac{1-\beta^N}{1-\beta}$, where $\beta = \sqrt\frac{t_{th}}{t_{th}+1/f}$ and the cooling time of the focal spot by thermal diffusion $\alpha$ is $t_{th} = l_{abs}^2/\alpha$ ($l_{abs}$ is the energy deposition depth taken equal to the thickness $h$ of Ta). The repetition rates of GHZ and MHz bursts are used together with the laser repetition rate. The dashed lines shows an expected effect with only one repetition rate (one corresponding time constant $1/f$); see discussion in the text. (b) Temperature rise inside single burst when $T_1 = 10^\circ$C and separation of pulselets is 0.4~ns (inside the 4~ns or 2.5~GHz first burst). The color-marked region is temperature rise by the first 10 pulses (close to a linear increase).         }
\end{figure*}

The brittle nature of nano-localized formation of \ce{Ta2O5} is evident from nano-cracks and holes present at the initial stages of ripple formation when single fs-pulses were used (SEM in Fig.~\ref{f-dipo} and Fig.~\ref{f-sem}). This is caused by the oxygen embitterment known in metals~\cite{Ma24}. In contrary, the BiB-mode provided controlled oxidation throughout the 200~nm film of Ta with sub-diffraction width along the scanned line (Fig.~\ref{f-step}). Importance of thermal conditions during oxidation for smooth nano-crack-free formation of micro-limes is further highlighted by comparison of fs vs. ps pulselets in the BiB mode of writing (Fig.~\ref{f-pico}). Very similar oxidized lines were made by 200~fs and 20~ps pulslets in the same $N_pN_n = 100$ BiB-mode writing at the same pulse energy and fluence, while the intensity differed by 100 times. Optical nonlinearity is less important, hence, the intensity in DLW oxidation of the metal surface. Considering the optical energy deposition into skin depth of few-tens-of-nanometers, and exothermal nature of oxidation, the high repetition rate and large density of ps- and ns-burst pulses are more important than high intensity. Importantly, when single 20~ps pulses at 602.7~kHz repetition rate were used to modify Ta 200 nm film, strong ablation occurred. Ablation was still absent, and oxidation was possible with single 1~ps and 5~ps pulses (not shown for brevity).    
\subsection{Heating in steps}

Heat diffusion is obviously an important factor for the oxidation of Ta. When polarization of a writing beam is perpendicular to the scan direction, an additional direct contribution of the electronic heat conductivity is added along the polarization (oscillating electrons) during the pulse duration. This was shown to be a contributing factor in ablation ripples formation~\cite{17sr39989}. Polarisation also defines direction for electron injection, is perpendicular to metal-oxide interfaces of newly formed oxide region and contributes to chemical modifications~\cite{11jnn2814}. 

\begin{figure*}[b!]
\centering\includegraphics[width=1\textwidth]{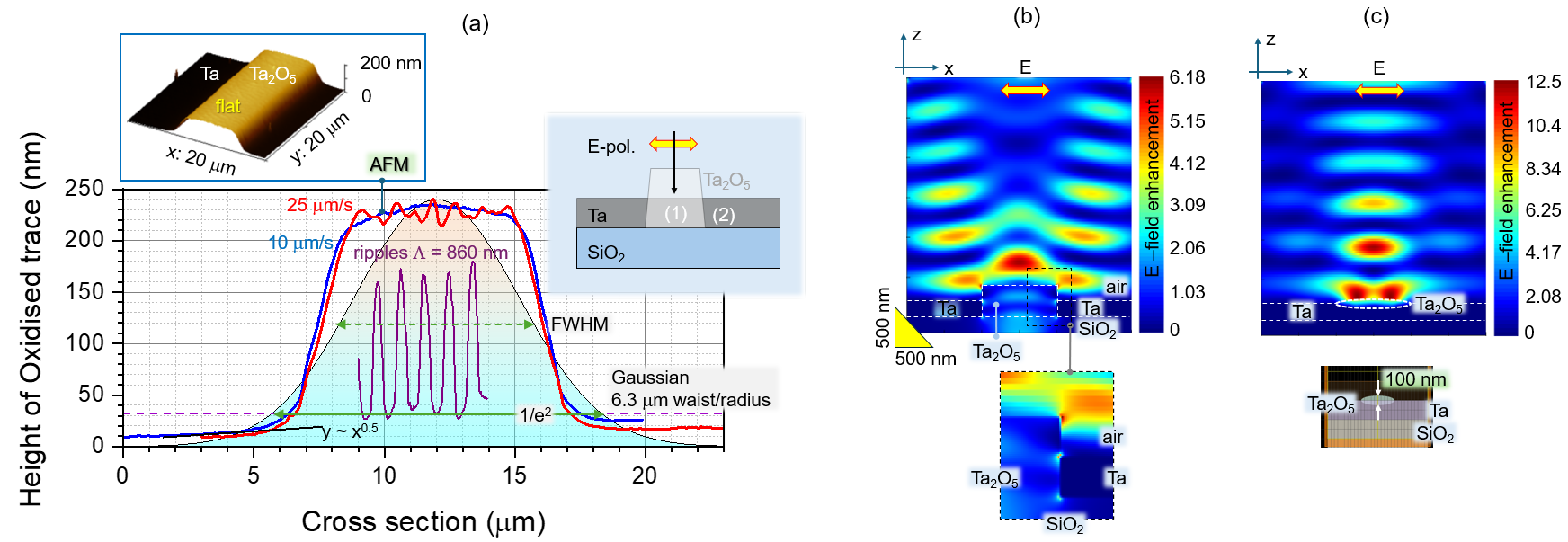}
\caption{\label{f-afm} (a) AFM cross sections of \ce{Ta2O5} lines written at 10 and 25~$\mu$m/s with $E_p = 1.35$~nJ in BiB-mode ($10\times 10$ ps-ns bursts); polarisation is linear (Fig.~\ref{f-step}(a)). Ripples' AFM cross section was measured perpendicular to the polarisation of the writing beam; pulse energy 0.38~nJ. Gaussian profile with waist (radius) $x_0 = 6.3~\mu$m is plotted as $y = 240e^{-2x^2/x_0^2}$; the FWHM is $\sqrt{2\ln2}x_0 = 7.4~\mu$m. The only region where diffusional height vs cross section scaling was applicable is shown with the $y\propto\sqrt{x}$ line (it was decided from the log-log plot). (b) FDTD simulations of light field enhancement at fully oxidised Ta film; the bottom-inet shows enhancement of normal $E^{(n)}$ component in the lower refractive index side of the interfaces. (c) Same as (b) only in partly oxidised Ta.      }
\end{figure*}

The enthalpy of Ta vaporization is 750~kJ/mol while Ta oxidation to \ce{Ta2O5} release $-~2038$~kJ/mol of energy at normal conditions~\cite{Holle}. This is a highly exothermic process, which provides a positive feedback loop to BiB fs-laser driven oxidation. With the 
specific heat capacity of Ta $c = 0.14$~J/(g$.$K), the energy required to reach efficient Ta oxidation temperature of $600^\circ$C ($\Delta T = 580$~K from RT) for a 1~mol of Ta is estimated as $Q=mc\Delta T\simeq 14.7$~kJ, where the molar mass of Ta $m = 180.95$~g/mol. This value is 138.6 times lower than the energy released in the production of the same amount (1~mol) of \ce{Ta2O5}. These estimates illustrate how oxidation of Ta was harnessed by the BiB mode of fs-laser writing. Melting temperature of Ta is 3017$^\circ$C and the latent heat of melting $\sim 35$~kJ/mol. 
To reach the melting point of Ta, 5.2 times larger amount of energy $\sim 76.4$~kJ would be required as compared to reaching 600$^\circ$C. Formation of molten material on top of oxidized region along laser tracks at the highest powers was apparent (Fig.~\ref{f-step}(b)) and can be provided partly by oxidation released heat as well as laser energy deposition. Temperature localisation and rise is also helped by a low thermal diffusivity of Ta $\alpha\simeq 20 
$~mm$^2$/s; for \ce{SiO2} it is 0.4~mm$^2$/s. 
Oxidation is a self-limiting process since the oxidized region becomes transparent and laser energy deposition stops. 

The qualitative mechanism of heating in a BiB-mode can be made using expression developed in continuous evaporation of targets for coatings when high-repetition tens-of-MHz rate ps-lasers are used~\cite{Gam}. If a temperature jump by the first pulse is $T_1$, the the temperature after sequence of $N$ pulses is $T_N = T_1\times\frac{1-\beta^N}{1-\beta}$, where $\beta = \sqrt\frac{t_{th}}{t_{th}+1/f}$ with the cooling time of the focal spot by thermal diffusion $\alpha$ is $t_{th} = l_{abs}^2/\alpha$; $f$ is the repetition rate. With separate ps- and ns-bursts at corresponding frequencies, the time constants are $1/f$: 4~ns, 16~ns, and 1660~ns for the main repetition rate of the fs-laser. The cooling time can be approximated by the diffusional cooling the thickness of Ta $h = 200$~nm. This assumption is justified by optical energy deposition depth which is close to $\lambda/4$. The result is shown in Fig.~\ref{f-temp}(a) where the first pulse is increasing surface temperature by 80$^\circ$C. Then saturation occurs very fast to the end of the 10-pulses burst. This is caused by fast cooling $t_{th} = 2$~ns while the GHz-bursts have time constant of $1/f_{GHz} = 4$~ns. From the saturated value (in this case $T_{10} = 235^\circ$C), heating repeats by MHz-burst only now with longer time constant $1/f = 16$~ns. Finally, the running repetition rate of laser corresponds to $1/f = 1660$~fs (even less of heat accumulation). The MHz-burst saturates temperature even faster than GHz-burst since $t_{th}\ll 1/f_{MHz}$. From saturation temperature of $T_{10} = 625^\circ$C, final rising step takes place. This toy model (Fig.~\ref{f-temp}(a)) captures the main tendency that fast saturation of temperature is taking place and small $N=10$ number of pulses in the burst is enough to reach saturated conditions. This is based on the cooling time being fast within the skin-depth of metal. What is also important, that if only one burst mode is used (or one main laser repetition rate), the temperature rises are smaller (dashed lines in Fig.~\ref{f-temp}(a)). 

This toy model use the repetition rates of the bursts, however, inside the burst there are 200~fs pulses separated by 0.4~ns. Those pulses produce the $80^\circ$C jump in the model (Fig.~\ref{f-temp}(a)) achieved by a cumulative action of 10 pulses with separation smaller than the cooling time $0.4$~ns$<t_{th}\equiv 2$~ns and $T_1 = 10^\circ$C (Fig.~\ref{f-temp}(b)). Hence, the thermal accumulations is the strongest during the first 10 pulses, which then are entering heating protocol simulated in Fig.~\ref{f-temp})(a).  

In real experiment, an exotermic heat of oxidation, melting of already oxidised top film of \ce{Ta2O5} (melting temperature $1870^\circ$C) all can change energy deposition conditions. The analysis shown in Fig.~\ref{f-temp} only serves as a qualitative guideline. However, the controlled and localised heating by BiB-mode is captured. 

\subsection{Near-field light intensity distribution at Ta-\ce{TaO5} interfaces}

The cross section of oxidised profile measured by AFM is shown in Fig.~\ref{f-afm}(a). Twice larger height as compared to the original Ta thickness of 200~nm signifies fully oxidised \ce{Ta2O5} as was determined from oxidation of the same sample in furnace. At such conditions even 2.5 times different exposure yielded in the same width of the oxide line. The width of the fully oxidised line is close to the full width at half maximum (FWHM) of the focal spot (assumed Gaussian). Once the Ta-\ce{Ta2O5} interface is formed with polarisation normal to it ((1)-(2) in the inset of Fig.~\ref{f-afm}), the boundary condition for the permittivity $D = \epsilon E$ across the interface is $D_{(1)} - D_{(2)} = \sigma_s$, where $\sigma_s$ is the surface charge density and permittivity $\epsilon = \Tilde{n}^2 = (n+i\kappa)^2$ defined by the complex refractive index. Finite differences time domain (FDTD) calculations were used to model light field enhancement at the Ta-\ce{Ta2O5}-air interfaces for different polarisations of incident plane wave (Fig.~\ref{f-afm}(b-c)). FDTD simulations provides the exact solution of Maxwell's equations for a particular 3D geometry of the pattern. The enhancement of E-field at the interfaces for the normal component at the side of lower refractive index is clearly visualized (see inset in Fig.~\ref{f-afm}(b)). The enhancement of E-field along polarisation of incident field is consistent with promotion of oxidation observed in ripples formation. Situation when Ta oxidation is not across the entire thickness of film (Fig.~\ref{f-afm}(c)), favors a high intensity region to be placed closer to the Ta (in \ce{Ta2O5}), which is expected to enhance oxidation. Particular sample geometry influence the actual light field enhancement geometry, however, the main features are captured by this qualitative model (the light field enhancement for polarisation parallel to the oxidised line is shown in Fig.~\ref{f-para}). The FDTD simulations are consistent with self-organisation and ripple pattern formation observed in Ta using 530~nm/200~fs pulses~\cite{26a}.    

\subsection{Energy considerations for oxidation vs ablation}

If is useful to compare energies required for electron removal from surface (in a chemical picture an oxidation) by laser, i.e. the electron work function $w_e$ of Ta is $\sim$4.3~eV, and the cohesive (binding) energy $\varepsilon_b$ or the latent heat of evaporation 750~kJ/mol, which is $\varepsilon_b/[N_ae] = 7.77$~eV/atom, where $N_a$ is the Avogadro number and $e$ is the charge of electron~\cite{Riffe}. For ablation, the fs-laser driven electron should exceed the energy of $w_e+\varepsilon_b\simeq 12$~eV/atom. For oxidation of Ta-to-\ce{Ta2O5}, two Ta neighboring atoms should be ionised, which requires only $\sim 8.6$~eV. Afterwards, a spontaneous oxidation takes place in the air environment. This shows that without ablation, oxidation of metal can proceed. Formation of structural defects, grain boundaries, etc., are locations were ionisation potential is reduced and this facilitates a larger energy deposition which leads to ablation (Fig.~\ref{f-abla}). 

It is noteworthy, that TaN and a range of different stoichiometric Ta-N compounds are also feasible to be formed on the surface and, e.g., TaN has only $1.3$~eV/atom cohesive energy. Since \ce{Ta2O5} has much larger cohesive energy and oxide formation on the surface wins against nitridation. Howver, Ta-N compounds can be formed in pure \ce{N2} environment including the solid solution of Ta and N which is thermodynamically stable~\cite{Wu}. 
Competition between evaporation, oxidation, ablation and controlled electron-ion removal~\cite{23n1796} should open new fs-laser machining beyond conventional ablation at true nanoscale $\sim 100$~nm feature sizes and resolution beyond the sub-diffraction limit $\leq 1~\mu$m as shown here. Interestingly, nanoscale defects and oxidation sites, which we can reproducibly generate on a metallic mirror surface of Ta, occurs at much lower pulse fluence as compared to the ablation threshold. This is very similar to the 650~nm NBOHC defect in pure \ce{SiO2}~\cite{Razvan}, which is induced at 0.1~TW/cm$^2$ fs-laser pulse intensity and is a pre-cursor for a step-wise absorption which leads to the ablation threshold under multiple exposure~\cite{Balling}.       

\section{Conclusions and Outlook}

Direct write oxidation of Ta nano-film by fs-laser pulses was demonstrated in two modes: 1) sub-diffraction limited oxidation through the entire film thickness at the micrometer scale using burst BiB-mode and 2) sub-wavelength, self-organized ripple type oxidation using raster scanning of single fs-pulses. Both modes of oxide patterning were achieved at high 0.6~MHz repetition rate of a fs-laser. The BiB with maximised number 10 of pulses in the ps- and ns-bursts was used for controlled oxidation throughout 200~nm Ta film. Very small single puslets energies $\sim 1$~nJ were used for direct write oxidation. Even 20~ps pulselets in the same BiB mode were delivering a very similar width of oxidized lines as compared with 200~fs pulses. The advantage of bust mode is clearly demonstrated in terms of ablation-free structural modification of Ta, since in the case of single pulses, ablation is taking place in most of the direct write conditions. Surface nanoscale fractures are clearly recognizable for single fs-pulse structuring when ripples are formed.   

Due to a wide range of possible valence states of Ta in a range of oxides, DLW with the BiB-mode of surface patterning of Ta for photo-catalytically active surfaces is a promising direction of applied research. Combined metallic conductivity and optical transparency of oxides (dielectrics or semiconductors) needs further investigation. Also, nano-films of different materials can be alloyed by fs-laser exposure~\cite{23m1917} and the BiB-mode used in this study is promising for on-surface synthesis and patterning of such nano-scale materials with a high spatial resolution.   

\small\begin{acknowledgments}
L.G. received funding from the Research Council of Lithuania (LMTLT), agreement No. S-LT-TW-24-9. S.J. acknowledges support via ARC DP240103231 grant. H-H.H. and S.J. are grateful for a research stay at the Laser Research Center, Vilnius University, in 2025. Fs-fab station at Swinburne's Nanolab was funded via MCN-Vic-ANFF infrastructure grant and opened for users in Dec. 2024.  
\end{acknowledgments}

\bibliography{aipsamp}

\appendix
\setcounter{figure}{0}\setcounter{equation}{0}
\setcounter{section}{0}\setcounter{equation}{0}
\makeatletter 
\renewcommand{\thefigure}{A\arabic{figure}}
\renewcommand{\theequation}{A\arabic{equation}}
\renewcommand{\thesection}{A\arabic{section}}
\section{Expansion of thickness due to oxidation of Ta}

\begin{figure*}[h!]
\centering\includegraphics[width=1\textwidth]{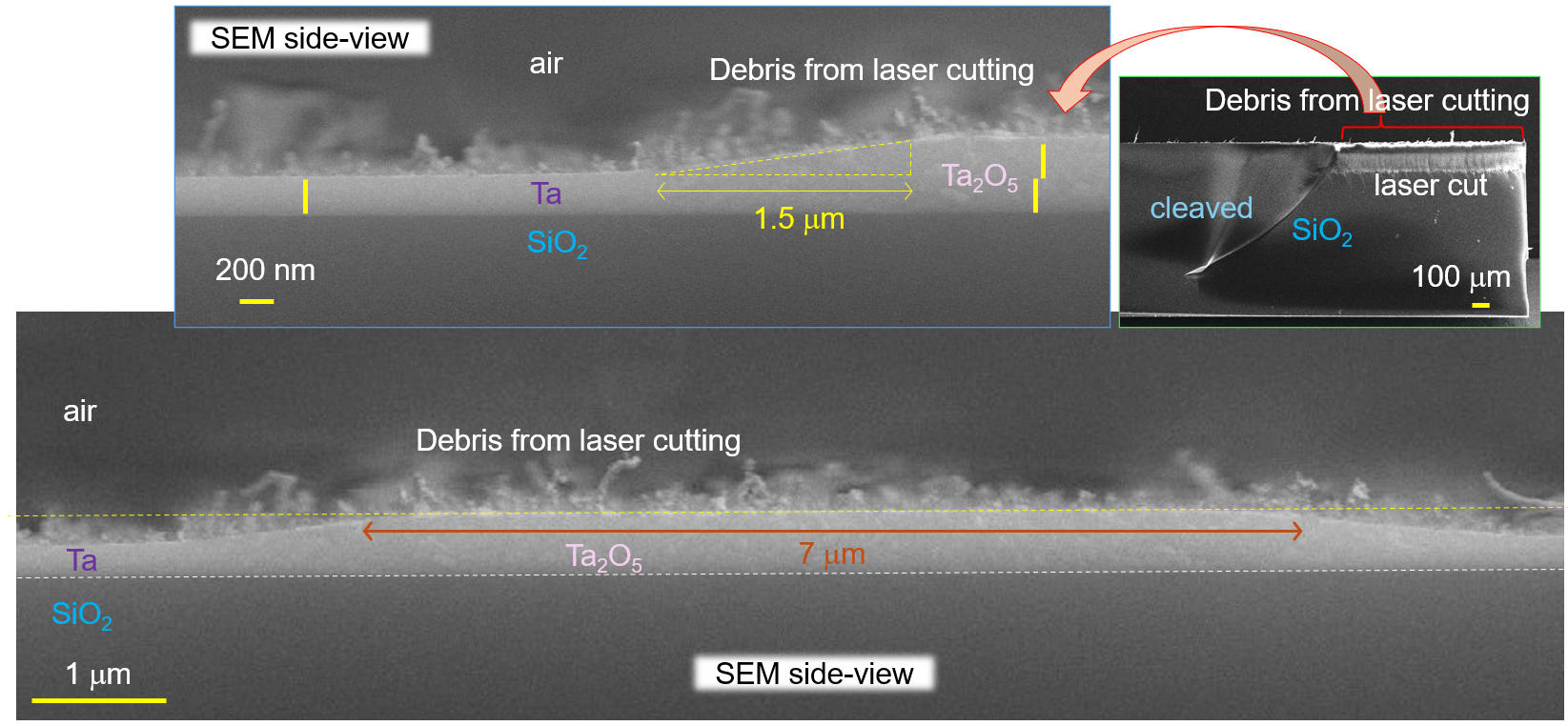}
\caption{\label{f-side} SEM side-view of Ta oxidized to \ce{Ta2O5} showing two times increased thickness; thickness measured from SEM: 240~nm Ta and 463~nm \ce{Ta2O5}. High-power laser ablation was used to inscribe a line with a section of not ablated segment for breaking (see, top-right inset). 
}
\end{figure*}

Figure~\ref{f-side} shows explicitly the Ta expansion during oxidation carried out in the BiB-mode. All the width of the oxidized region is smaller than the focal spot diameter $\sim 12.6~\mu$m. The top surface of flat. Side ramps of the transitional Ta-to-\ce{Ta2O5} region are within $1.5-1.7~\mu$m at the used specific exposure conditions. Those ramps should have partially oxidized Ta according to the local temperature; a $600^\circ$C is required for oxidation of Ta in air. The same origin of heat accumulation is recognizable at the beginning of laser modified lines for fs- and ps-BiB modes when single pulse energies are $\sim 1$~nJ (Fig.~\ref{f-pico}). 

It is noteworthy that the interface \ce{SiO2}-Ta is in a pristine state after \ce{Ta2O5} formation without delamination, crack initiation, or height modulation due to melting and material reflow. This is a key advantage for the application of the direct write oxidation of Ta. 

\section{Picosecond pulses in the BiB-mode of Ta oxidation}
\begin{figure*}[h!]
\centering\includegraphics[width=1\textwidth]{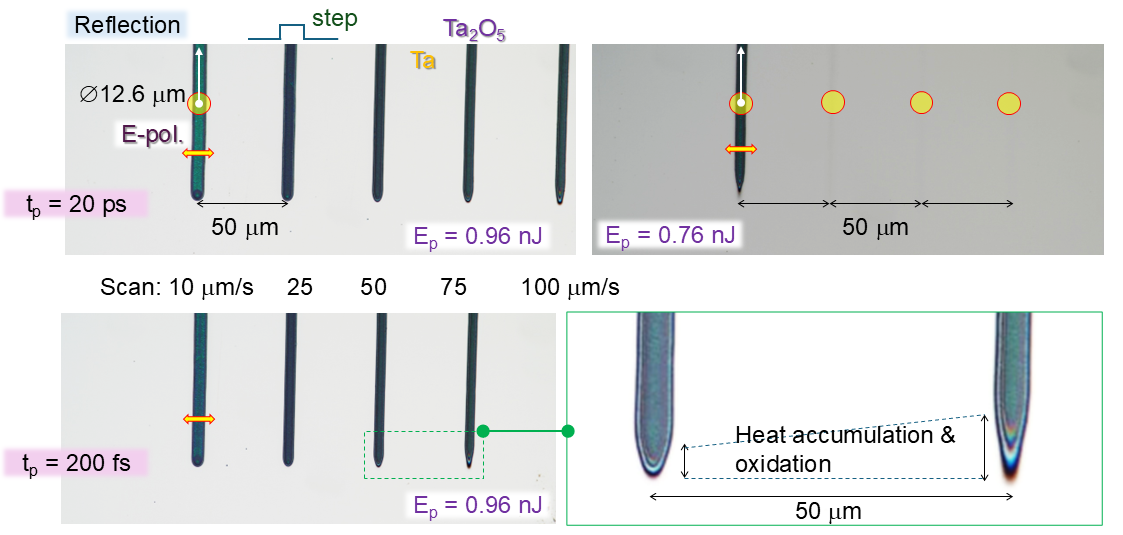}
\caption{\label{f-pico} Picosecond $t_p=20$~ps BiB-mode. Single pulselet energies were $E_p = 0.96$~nJ and 0.76~nJ. For comparison, 100-times shorter pulses $t_p=200$~fs BiB-mode oxidation of Ta at the same pulselet energy $E_f = 0.96$~nJ is shown (bottom row). The fluence per pulselet in 200~fs and 20~ps case is the same $F_f\equiv F_p = 0.77$~mJ/cm$^2$ for 0.96~nJ energy. The corresponding average intensities are very different $I_f = 3.85$~GW/cm$^2$ and $I_p = 38.5$~MW/cm$^2$ (by a factor of 100). Heat accumulation and oxidation until the steady state conditions are reached is shown in the close up view in the right-inset.  }
\end{figure*}
\begin{figure*}[h!]
\centering\includegraphics[width=0.9\textwidth]{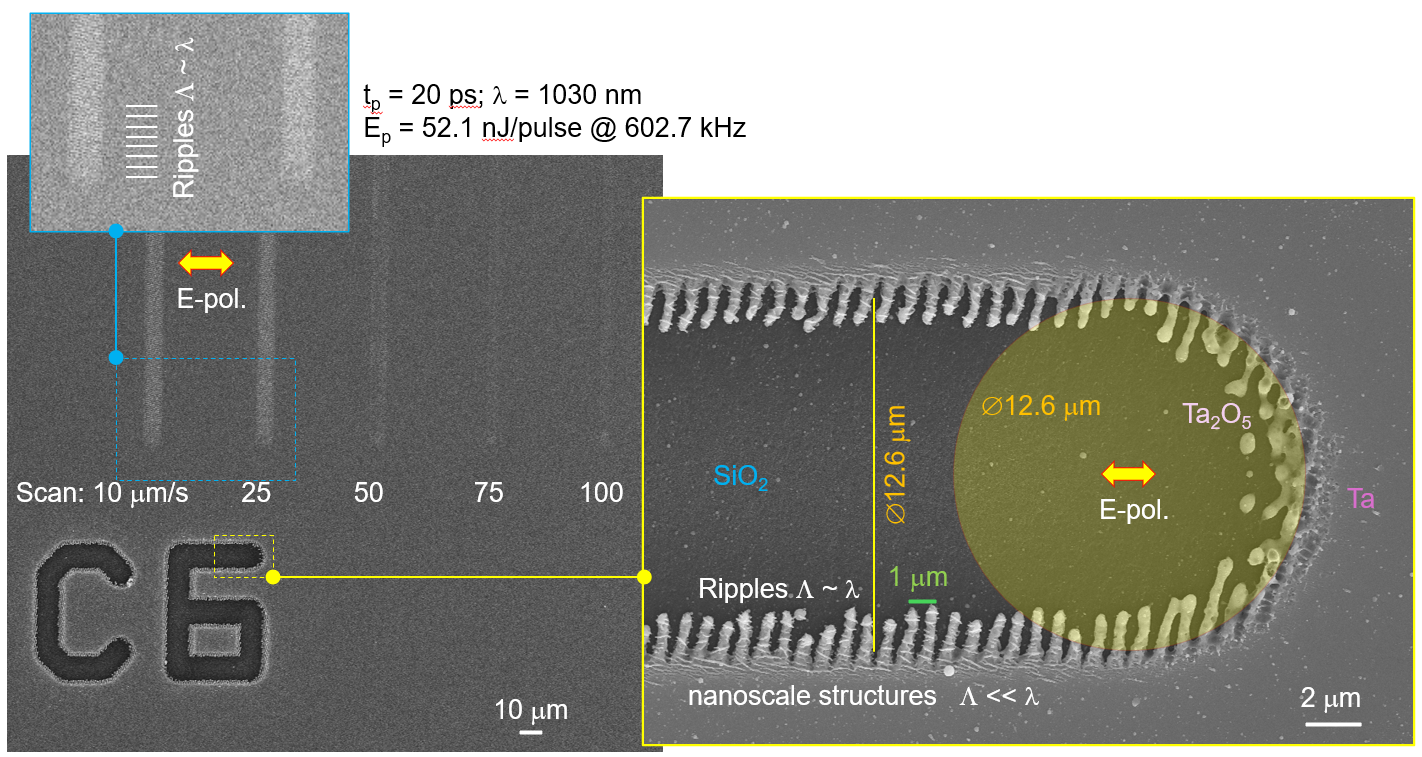}
\caption{\label{f-20ps} SEM images of Ta oxidized/ablated by 20~ps pulses at 602.7~kHz repetition rate with single pulse energy of 52.1~nJ. Conditions for the marker letter ``6'': it was recorded at the same pulse energy only with a larger pulse to pulse overlap; the position marking the focal spot is the starting point of the scan.}   
\end{figure*}
\begin{figure*}[h!]
\centering\includegraphics[width=0.6\textwidth]{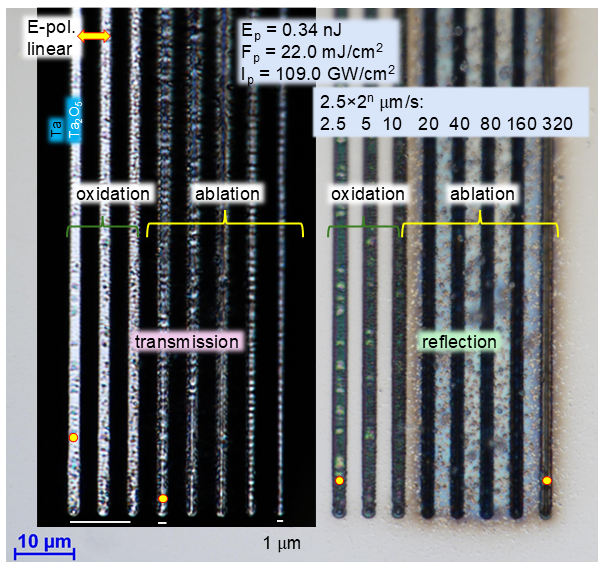}
\caption{\label{f-abla} Transmission (left) and reflection (right) images of laser oxidation/ablation under $NA = 0.9$ focusing.}   
\end{figure*}
\begin{figure*}[h!]
\centering\includegraphics[width=0.9\textwidth]{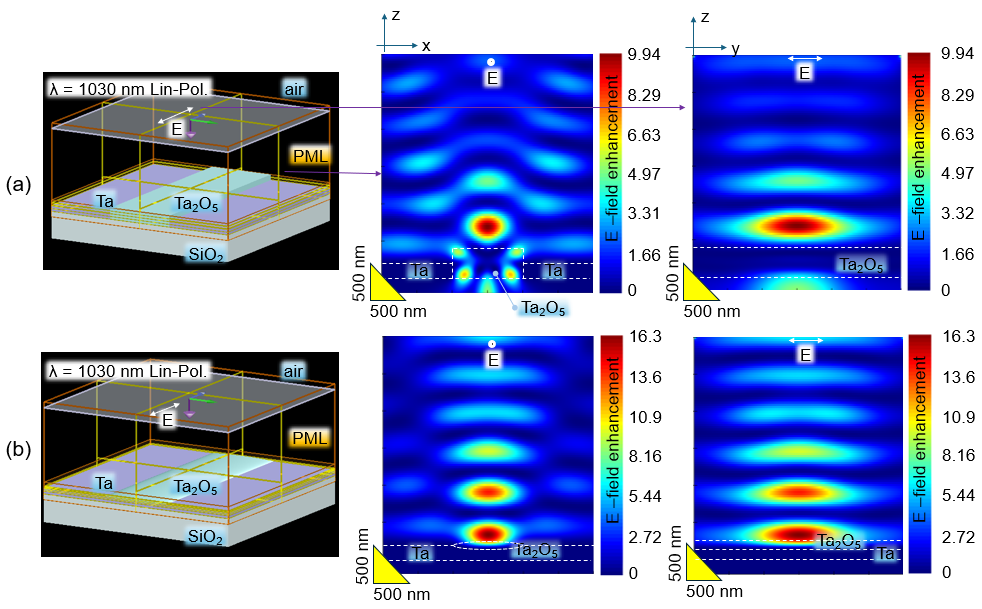}
\caption{\label{f-para} FDTD simulation (Lumerical solver, Ansys) of transverse cross section of the fully (a) and partly (b) oxidised Ta film when incident plane wave is polarised $E_y$ along the line. The two monitors are capturing light intensity with E-field parallel and perpendicular to the monitor. }   
\end{figure*}

Figure~\ref{f-pico} shows 20~ps BiB-mode oxidations, which are slightly more efficient as compared with 200~ps BiB-mode. There is the same fluence per single shortest pulse in both cases, however, the difference in average intensities is $10^2$ times.  

Interestingly, with pulses chirp-tuned to 20~ps, there were similar oxidized ripples' formation along the orientation of linear E-field polarization. However, larger intensities or slower scanning caused ablation when a single pulse rather than BiB mode was used. The ablated marker region shows a large exposure dose at the same pulse energy as used for the line pattern. Usual ripples with period $\Lambda\simeq 800$~nm are formed at the rim of the focal region when the entire Ta film of 200~nm was ablated.  

\section{From oxidation to ablation}\label{abla}

At the most tight focusing with $NA = 0.9$ objective lens, oxidation is achieved for the slowest scan speeds. Once scanning speed exceeds some value (dependent on pulse energy), e.g., 20~$\mu$m/s for 0.34~nJ/pulse, ablation becomes evident by formation larger debris fields (Fig.~\ref{f-abla}). The ablation of the entire Ta thickness of 200`nm was achieved and in transmission image the width of the line $\sim 1~\mu$m-wide smaller than the focal diameter of $1.4~\mu$m was observed. The pulse fluence was $22$~mJ/cm$^2$, which is typical for metal ablation. Apparently, when scanning is slow and heat accumulation is taking place in the BiB-mode, heating and oxidation is the most probable. At high scan speed, brittle fracture occurs. When single pulselet energy $E_p\simeq 0.1 - 0.2$~nJ and fluence below a single pulse ablation threshold for metal, apparently, the heat deposition and oxidation can be controlled without debris generation. This could be consistent with ionisation of metal, however, without energy required to overcome the binding energy of Ta. As a result, ionised surface is chemically active (a dangling bond) and can be readily oxidised via exothermic process.    

\end{document}